# Rigorous Theory of Optical Trapping by an Optical Vortex Beam

Jack Ng,[1] Zhifang Lin,[1,2] and C. T. Chan[1]

[1]*Department of Physics and William Mong Institute of Nano Science & Technology,*

*Hong Kong University of Science and Technology,*

*Clear Water Bay, Hong Kong, China.*

[2]*Department of Physics, Fudan University, Shanghai, China.*

**Abstract**

We propose a rigorous theory for the optical trapping by optical vortices, which is emerging as an important tool to trap mesoscopic particles. The common perception is that the trapping is solely due to the gradient force, and may be characterized by three real force constants. However, we show that the optical vortex trap can exhibit complex force constants, implying that the trapping must be stabilized by ambient damping. At different damping levels, particle shows remarkably different dynamics, such as stable trapping, periodic and aperiodic orbital motions.

Optical tweezers is a powerful tool for trapping mesoscopic objects. Applications range from the trapping and cooling of atoms, to large molecules such as DNA, and to microscopic particles and biological objects. As new methods to create beam profiles are being introduced, more exotic beams are used to trap particles and many of them, known as optical vortex (OV), carry angular momentum (AM). We will show that although the conventional stiffness constant approach works for a Gaussian beam, the theory of trapping by an OV is more complex and interesting. In the conventional approach, optical traps are usually characterized by three stiffness constants along the three principal axes, which are usually taken to be the Cartesian axes (e.g., *x*-polarized,

*z*-propagating beam*)*. Although such approach is pretty accurate in describing the ordinary optical tweezers, a more rigorous treatment reveals that the principal axes are not necessarily given by the Cartesian axes and, for an OV, the principal axes are not even "real". The rigorous theoretical procedure to obtain the principal axes is to diagonalize the force constant matrix $K_{ij} = \partial f_{light,i} / \partial x_j$ at equilibrium, where $f_{light,i}$ and $x_i$ are, respectively, the *i*-th Cartesian component of the optical force and particle displacement away from the equilibrium position. It is the eigenvalues of the force constant matrix that give the eigen force constants (EFC, or trap stiffness), while the eigenmodes determine the principal axes. We shall apply the force constant matrix formalism to study OV trapping [1,2,3,4,5,6,7,8,9,10,11,12,13,14,15,16,17,18,19,20,21,22,23,24], and we see that the difference between the conventional approach and the rigorous treatment is a qualitative one, to the extent that the EFCs can be complex numbers and we have to abandon concepts such as parabolic potential for the transverse directions. By analyzing the stability and simulating the dynamics of a particle trapped by OVs, it is found that the trapping stability of OVs generally depends on the ambient damping. In particular, in the presence of AM, the optical trapping may exhibit a fascinating variety of phenomena ranging from "opto-hydrodynamic" trapping (where the trapping is stabilized by the ambient damping) to supercritical Hopf bifurcation (where a periodic orbit is created as ambient damping decreases).

To illustrate the basic idea, let us starts from the linear stability analysis. Consider a trapped particle, near an equilibrium trapping (zero-force) position, the optical and damping forces [25] are $\vec{F} = m\, d^2\Delta \vec{x}/dt^2 \approx \ddot{\vec{K}}\Delta\vec{x} - \gamma\, d\Delta\vec{x}/dt$, where *m* is the mass of the particle, $\Delta\vec{x}$ is the particle's displacement from the equilibrium, $\ddot{\vec{K}}\Delta\vec{x}$ is the optical force, and $\gamma$ is the ambient damping constant. The eigenvalues $K_i$'s of the force matrix $\ddot{K}$ are precisely the EFCs and the eigenvectors of $\ddot{K}$ are the eigenmodes. For a trapping beam propagating along $\hat{z}$, the force constant matrix has the general form

$$\ddot{K} = \begin{bmatrix} a & d & 0 \\ g & b & 0 \\ e & f & c \end{bmatrix}, \tag{1}$$

where all elements are real numbers. The elements $\partial f_{light,x} / \partial z$ and $\partial f_{light,y} / \partial z$ are zeros by symmetry, because there is no induced force along the transverse plane as the particle is displaced along $\hat{z}$. The diagonal elements *a*, *b*, and *c* characterize three restoring forces, which are usually taken as the three stiffness constant in conventional approach. Off diagonal elements *d* and *g* characterize the rotational torques: i.e. as the particle is displaced along *x* (*y*), it experiences a torque that manifests as a force along the *y* (*x*) direction. As an OV carries orbital AM, the energy of the beam propagates along a helical path. There is a rotating energy flux in the transverse plane that would exert a torque on the particle, implying non-zero *d* and *g*. On the other hand, $d = g = 0$ for beams that carry no AM.

By diagonalizing Eq. (1), we obtain three EFCs: $K_{axial} = c$ and $K_{trans\pm} = \left[ a + b \pm \sqrt{(a-b)^2 + 4dg} \right] / 2$. Conservative mechanical systems can be described by a potential energy *U*, and their force constant matrix is a real symmetric matrix (i.e. $K_{ij} = -\partial^2 U / \partial x_i \partial x_j = K_{ji}$), which then follows that all eigenvalues are real. However, optical force is non-conservative as the particle can exchange energy with the beam, and its $\ddot{K}$ is in general a real but non-symmetric matrix. Consequently, its eigenvalues can carry conjugate pair of complex numbers. It can be seen that when

$$-4dg > (a-b)^2, \tag{2}$$

$K_{trans\pm}$ are a conjugate pair of complex numbers. Thus, in general, an optically

trapped particle may not be characterized by three real force constants if the beam carries AM. For beams that have no AM, $d = g = 0$, all the EFCs are real numbers. On the other hand, an OV beam, well known for its AM carrying characteristic, leads to non zero *d* and *g*, and thus (2) can be fulfilled under certain conditions. The existence of complex EFCs implies that the common notion of a parabolic potential in an optical trap is no longer meaningful. Whether complex EFC occurs depend on the competition between the beam asymmetry and the AM. Equation (2) cannot be fulfilled when $|a-b|$ is large. Since *a* (*b*) is the restoring force constant for the *x*- (*y*-) direction, a large $|a-b|$ implies a large asymmetry between the two coordinate axes. The underlining physics is that for large asymmetry, the beam can pin a particle to one of its axis, preventing it from "falling" into the OV. On the contrary, if there is weak or no asymmetry ($|a-b|\simeq 0$), the trapped particle is not tied to the coordinate axis, and will thus be swiped by the OV. As a result, its AM and energy will accumulate. If there is no dissipation ($\gamma = 0$) the particle will orbit around the beam center with increasing speed and eventually escape from the trap [26]. It is therefore concluded that in general the OV trapping cannot be achieved solely by light. It is the dissipation in the suspending medium that keeps the particle's kinetic energy and AM bounded, rendering the particle trapped [26]. Such a state of OV trapping is believed to be what the experiments observed, instead of the pure gradient force trapping. For the case of circularly polarized beams (LG or Gaussian), cylindrical symmetry mandates that $|a-b|=0$, implying that the transverse EFCs are always complex. This means that circularly polarized beam cannot trap without dissipation. More mathematical details can be found in on-line materials [34].

We now proceed to show concrete examples in which the EFCs are indeed complex. We model the incident trapping beam by using the highly accurate generalized vector Debye integral [27,28], where the focusing of the incident laser beam by the high numerical aperture (N.A.) object lens is treated using geometrical optics, and then the focused field near the focal region is obtained using the angular spectrum representations. The use of geometrical optics in beam focusing is fully justified as the lens is macroscopic in size, and all remaining parts of our theory employs classical electromagnetic optics. With the strongly focused beam given by the vector Debye integral, the Mie theory is then applied to calculate the scattered field, and then the Maxwell stress tensor formalism is applied to compute the optical force. [29,30] We note that the formalisms we use have been proven to agree well with experiments [28,31,32,33]. Fig. 1(**a**) shows $K_{trans\pm}$ for an LG beam focused by a high N.A. water immersion objective. The beam is linearly polarized with wavelength $\lambda = 1064$ nm, topological charge $l$=1, N.A.=1.2, and filling factor $f$=1. The trapping is in water ($\varepsilon_{water} = 1.33^2$) and the sphere is made of polystyrene ($\varepsilon_{sphere} = 1.57^2$ and mass density $\rho = 1050$ kg m$^{-3}$). The axial EFC is not plotted, as it is always real and negative, indicating that the particle can be trapped along the axial direction solely by gradient forces. It can be clearly seen from Fig. 1(a) that at certain ranges of particle sizes, $\text{Im}\{K_{trans\pm}\} \neq 0$, and this numerical results manifest the assertion that the OV trap cannot be characterized by three real force constants in general. At these particle sizes, $K_{trans+} = K_{trans-}^*$, and the two curves corresponding to $\text{Re}\{K_{trans\pm}\}$ merge together. Note that only the absolute value of $\text{Im}\{K_{trans\pm}\}$ is plotted in Fig. 1(a). When the EFCs are all real numbers, the behavior of the trapped particle is qualitatively similar to that of the ordinary optical trapping by conventional optical tweezers. This corresponds to the scenario that either the beam's AM is weak (small $d$ and $g$), or the asymmetry of the beam ($|a-b|$) is large. We note that the existence of region where $\text{Im}\{K_{trans\pm}\} \neq 0$ implies that in low viscosity media, only particles of certain sizes can be trapped. For complex EFCs, the eigenmodes

corresponding to the complex EFCs are [34]:

$$\Delta \vec{x}_{\pm}(t) = A_{\pm} e^{-\text{Im}(\Omega_{\pm})t} \left\{ \begin{array}{l} \text{Re}(\vec{V})\sin[\text{Re}(\Omega_{\pm})t + \phi_{\pm}] \\ + \text{Im}(\vec{V})\cos[\text{Re}(\Omega_{\pm})t + \phi_{\pm}] \end{array} \right\}, \quad (3)$$

where $\vec{V}$ is the eigenvector corresponding to eigenvalue $K_i$ of the force constant matrix $\vec{\vec{K}}$. $\{A_{\pm}, \phi_{\pm}\}$ are to be determined from initial conditions, and

$$\text{Re}(\Omega_{\pm}) = \mp(\Delta_R^2 + \Delta_I^2)^{1/4} \sin(\delta/2)/2m,$$
$$\text{Im}(\Omega_{\pm}) = \gamma \pm (\Delta_R^2 + \Delta_I^2)^{1/4} \cos(\delta/2)/2m, \quad (4)$$

$$\delta = \begin{cases} \tan^{-1}(\Delta_I/\Delta_R), & \text{if } \Delta_R > 0, \\ \pi - \tan^{-1}(\Delta_I/|\Delta_R|), & \text{if } \Delta_R < 0, \end{cases} \quad (5)$$

where $\Delta_I = 4m\,\text{Im}\{K_i\}$ and $\Delta_R = \gamma^2 + 4m\,\text{Re}\{K_i\}$. The modes are stable if and only if both $\text{Im}(\Omega_{\pm}) > 0$. If $\text{Re}\{K_i\} > 0$, one of the two modes is always unstable such that upon small perturbation, the particle will spiral outward and leave the trap. If $\text{Re}\{K_i\} < 0$, the mode is unstable for $\gamma < \gamma_{critical} = \sqrt{m}\,|\text{Im}\{K_i\}|/\sqrt{|\text{Re}\{K_i\}|}$. However this equilibrium can be stabilized by increasing $\gamma$ to beyond $\gamma_{critical}$. We label this kind of mode as quasistable modes, in which the stability of the modes depends on the ambient damping. The complex modes described by Eq. (3) correspond to spiral motions, which means that the particle is absorbing AM from the beam. The converse is also true: these spiral modes can exist only when the particle can absorb AM.

We note in Fig. 1 that in our specific example, $\text{Re}\{K_{trans\pm}\} > 0$ for sphere radius $R < 0.36\,\mu m$ (radius of the intensity ring $\sim 0.33\,\mu m$, see Fig. 2(a)), which means that small dielectric particles are unstable, as reported in experiments [14,35]. The small dielectric particles are attracted by the high intensity ring and under sufficient damping, these small particles will orbit along the ring [5]. On the other hand, $\text{Re}\{K_i\} < 0$ for $R > 0.36 \mu m$. The sphere is bigger than the intensity ring, so that the gradient force drives the sphere to the beam center. A phase diagram for the optically trapped particle is given in Fig. 1(b). At $a = 0.36 \mu m$, $\gamma_{critical} \to \infty$ as $\text{Re}\{K_{trans\pm}\} \to 0$. The equilibrium point at $(x, y) = (0, 0)$ is unstable for $a < 0.36 \mu m$ at any values of damping. For $a > 0.36 \mu m$, the white (shaded) region where $\gamma > \gamma_{critical}$ ($\gamma < \gamma_{critical}$) is the regime where the damping is sufficient (insufficient) to stabilize the particle. We note that when the EFC is real, no damping is required for stability since $\gamma_{critical} = 0$. According to Stoke's law, the damping constant of water and air are, respectively, $1.9 \times 10^4 R$ (pN $\mu s/\mu m^2$) and $3.3 \times 10^2 R$ (pN $\mu s/\mu m^2$). The damping of water is much larger than $\gamma_{critical}$ plotted in Fig. 1(b), and thus unless one uses high laser power, one shall observe stable trapping in water, in agreement with existing experiments. The damping of air is of the same order of magnitude of $\gamma_{critical}$ plotted in Fig. 1(b), consequently for an experiment conducted in air, one shall be able to see the transition between the stable and unstable state, depending on the laser power employed.

It is now clear that a particle trapped by an OV is stable if $\gamma > \gamma_{critical}$ and unstable if $\gamma < \gamma_{critical}$. Nevertheless, the case of $\gamma \approx \gamma_{critical}$ is non-hyperbolic (the linear term vanishes), and thus the higher order terms are important. In that case, we numerically integrate the full equation of motion $m \, d^2 \Delta \bar{x}/dt^2 = \bar{f}_{light} - \gamma \, d\Delta \bar{x}/dt$, using an adaptive time-step Runge-Kutta-Verner algorithm [29]. Fig. 2(a) shows the field intensity on the focal plane for a right circularly polarized LG beam, with a dark central spot and a high intensity ring of radius ~$0.33 \mu m$. Fig. 2(b)-(f) show the trajectories of a 1-$\mu m$-diameter particle illuminated by the LG beam (power=550 mW), in the order of decreasing damping. When there is strong damping, as shown in

Fig. 2(b) where $\gamma = 550$ pN $\mu$s/$\mu$m, the trapped sphere exhibits damped oscillation upon small perturbation and settles into a stable equilibrium position. For weaker damping, the sphere initially spirals outward, and then settles into a periodic circular orbit (see Fig. 2(c) where $\gamma = 110$ pN $\mu$s/$\mu$m ). Such bifurcation of a stable equilibrium into an unstable equilibrium and a stable periodic orbit is known as a supercritical Hopf bifurcation [29]. If we further reduce the damping, the radius of the circular orbit increases, as shown in Fig. 2(d) where $\gamma = 55$ pN $\mu$s/$\mu$m. If the damping decreases further, the particle goes into an exotic orbit around the intensity ring, as shown in Fig. 2(e) where $\gamma = 5.5$ pN $\mu$s/$\mu$m. When there is no damping (see Fig. 2(f), $\gamma = 0$ pN $\mu$s/$\mu$m ), the particle initially fluctuates around the equilibrium with increasing amplitude, and eventually escapes from the trap due to the accumulation of AM. If a small imaginary part is introduced into the dielectric constant of the particle, the introduced absorption will compete with the light scattering, reducing the amount of AM that is transferred to the orbital motion of the particle, and the particle will now spin along its own axis. In other words, small absorption may in fact favor the transverse trapping, though it degrades the axial trapping.

Our analysis reveals that for an AM carrying beam, its EFCs can be complex numbers. In the case of complex EFCs, when there is sufficient (insufficient) damping a particle can (cannot) be stably trapped. There is an intermediate range of damping in which the particle will be driven into exotic periodic or aperiodic orbital motions. Finally, we note that as the ambient damping force plays an important role in the OV trapping, it should be more accurately termed "opto-hydrodynamic trapping".

We have also applied the stability analysis to other types of focused beams [34] with different N.A., and we find that we can observe complex EFCs whenever the beam carries AM.

This work is supported by Hong Kong RGC grant 600308. ZFL was supported by NSFC (Grant number 10774028), PCSIRT, and MOE of China (B06011). Jack Ng was partly supported by NSFC (Grant number 10774028).

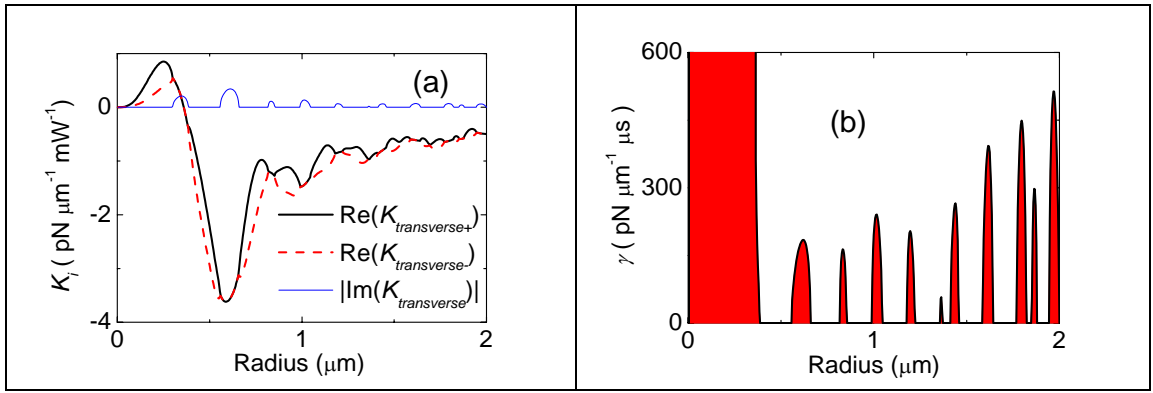

Fig. 1 The incident beam is a linearly polarized LG beam with $\lambda = 1064$ nm, $l=1$, $f=1$, and N.A.= 1.2. **(a)** The transverse EFCs. **(b)** Phase diagram for a particle trapped at a power of 1W. The white (red) regions are unstable (stable). The black line marks $\gamma_{critical}$.

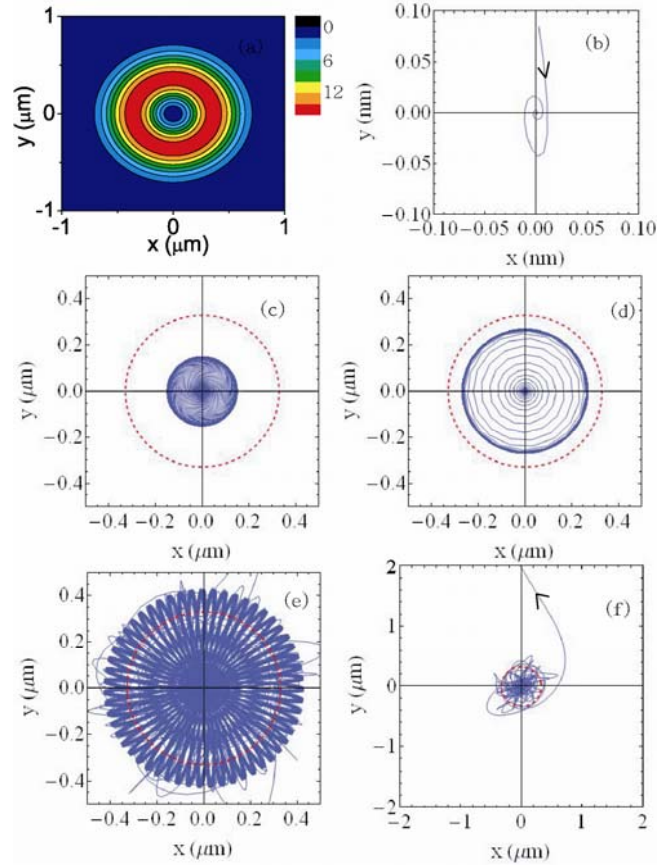

Fig. 2 **(a)** The focal plane intensity (arbitrary units) of a right polarized LG beam with $\lambda = 1064$ nm, $l=1$, $f=1$, and N.A.=1.2. **(b)-(f)** The trajectory (blue) of a 1-micron-diameter particle trapped by a 550 mW beam. The red dotted lines are the approximate radius of the intensity ring of the trapping beam. The damping constants $\gamma$ for each panel, in unit of pN $\mu$s/$\mu$m, are given by **(b)** 550, **(c)** 110, **(d)** 55, **(e)** 5.5, and **(f)** 0. The arrows in **(b)** and **(f)** indicate the direction of motion.

# Appendix I: Linear Stability Analysis

In this appendix, we give more details about the formalism on the linear stability analysis for a particle trapped by an arbitrary incident light beam.

**A. Linearized equation of motion**

We denote the displacement of the particle away from the equilibrium position by the position vector $\Delta \vec{x} = (\Delta x, \Delta y, \Delta z)$. The equation of motion of the particles are given by

$$m\frac{d^2 \Delta \vec{x}}{dt^2} = \vec{f}_{light}(\Delta \vec{x}) - \gamma \frac{d\Delta \vec{x}}{dt}, \tag{I.1}$$

where $m$ is the mass of the particle, $\vec{f}_{light}(\Delta \vec{x})$ is the optical force, $\gamma$ is the damping constant for the particle in the suspending medium. The frictional term in (I.1) is added to account for the Stoke's drag between the particle and the suspending medium, and we have deliberately neglected the Brownian term in (I.1), as it is of significance only at low laser power. If the displacement $\Delta \vec{x}$ is small compare to the wavelength of incident light ($|\Delta \vec{x}| \ll \lambda$), it is possible to simplify (I.1) with a linear approximation with respect to the displacement. The linearized equation of motion is

$$m\frac{d^2 \Delta \vec{x}}{dt^2} \approx \ddot{\vec{K}} \Delta \vec{x} - \gamma \frac{d\Delta \vec{x}}{dt}, \tag{I.2}$$

where

$$(\ddot{\vec{K}})_{jk} = \frac{\partial (\vec{f}_{light})_j}{\partial \Delta x_k}\bigg|_{\Delta \vec{x}=\vec{0}} \tag{I.3}$$

is the force constant matrix. We note that the zero-th order term in (I.2) vanishes, because we are expanding the force near an equilibrium where $\vec{f}_{light}(\Delta \vec{x} = \vec{0}) = \vec{0}$.

Introducing the transformation

$$\Delta \vec{x} = \ddot{\vec{V}} \vec{\eta}, \tag{I.4}$$

where $\eta_i$'s are the normal coordinates, and the columns of $\ddot{\vec{V}}$ are the eigenvectors

of $\vec{\vec{K}}$ so that $\vec{\vec{K}}$ is diagonalized with eigenvalues $K_i$'s:

$$\vec{\vec{V}}^{-1}\vec{\vec{K}}\vec{\vec{V}} = \sum_i \hat{x}_i \hat{x}_i^T K_i. \tag{I.5}$$

Here $\hat{x}_i$ is the unit vector along the Cartesian axes. In a conservative mechanical system that can be described by a potential energy $U$, the force constant matrix is symmetric (i.e. $K_{ij} = -\partial^2 U / \partial x_i \partial x_j = K_{ji}$) and can only give real negative or real positive eigen force constants. However, optical force is non-conservative, and $\vec{\vec{K}}$ is in general a real valued non-symmetric matrix (i.e. $K_{ij} \neq K_{ji}$). As such, its eigenvalues and their corresponding eigenvectors, can be real numbers or a conjugate pair of complex numbers. After substituting (I.4) into (I.2), the equation of motion is now decoupled into three independent equations:

$$m\frac{d^2\eta_i}{dt^2} = K_i\eta_i - \gamma\frac{d\eta_i}{dt}. \tag{I.6}$$

Equation (I.6) is a second order linear ordinary differential equation, which may be solved by the standard technique of substituting

$$\eta_i = \eta_{0i} e^{i\Omega_i t}. \tag{I.7}$$

where $\eta_{0i}$ and $\Omega_i$ are independent of time. It turns out that the solution to (I.6) can be categorized according to the eigenvalues $K_i$ or equivalently the natural frequency of the eigenmode defined as

$$\Omega_{i0} = \sqrt{-K_i/m}. \tag{I.8}$$

### B. Types of Eigenmodes

#### 1. Unstable mode characterized by an imaginary natural frequency

If $K_i$ is real and positive, the corresponding natural frequency is purely imaginary and the corresponding solution is

$$\Delta \vec{x}_i(t) = e^{-\gamma t/2m} \vec{V}_i \left[ A_i e^{-\sqrt{(\gamma/2m)^2 + |\Omega_{i0}|^2} t} + B_i e^{\sqrt{(\gamma/2m)^2 + |\Omega_{i0}|^2} t} \right], \tag{I.9}$$

where $A_i$ and $B_i$ are unknown constants to be determined from the initial conditions. The mode is unstable because (I.9) diverges with time.

#### 2. Stable mode characterized by real natural frequency

If $K_i$ is real and negative, the natural frequency is purely real and the motions of the particles are that of a damped harmonic oscillator. For $(\gamma/2m)^2 > \Omega_{i0}^2$,

$$\Delta \vec{x}_i(t) = e^{-\gamma t/2m} \vec{V}_i \left[ A_i e^{-\sqrt{(\gamma/2m)^2 - \Omega_{i0}^2} t} + B_i e^{\sqrt{(\gamma/2m)^2 - \Omega_{i0}^2} t} \right], \tag{I.10}$$

where $A_i$ and $B_i$ are unknown constants to be determined from initial conditions. The oscillation is over damped. For $(\gamma/2m)^2 = \Omega_{i0}^2$,

$$\Delta \vec{x}_i(t) = e^{-\gamma t/2m} \vec{V}_i \left[ A_i + B_i t \right], \tag{I.11}$$

where $A_i$ and $B_i$ are unknown constants to be determined from initial conditions. The oscillation is critically damped. For $(\gamma/2m)^2 < \Omega_{i0}^2$,

$$\Delta \vec{x}_i(t) = A_i e^{-\gamma t/2m} \vec{V}_i \sin\left[ \sqrt{\Omega_{i0}^2 - (\gamma/2m)^2} t + \phi_i \right], \tag{I.12}$$

where $A_i$ and $\phi_i$ are unknown constants to be determined from initial conditions. The oscillation is under damped.

    The trajectories of the solutions (I.10), (I.11) and (I.12) are all bounded as time increases, accordingly they are all stable.

#### 3. Complex mode characterized by a complex natural frequency

As the force constant matrix is non-symmetric, a complex conjugate pair of eigenvalues can occur. To obtain the trajectories associated with the conjugate pair of eigenvalues $K_i$ and $K_i^*$, it suffices to consider only $K_i$ where Im{ $K_i$ }>0. The solutions associated with $K_i^*$ are the same as that of $K_i$. The solutions are

$$\Delta \vec{x}_{i+}(t) = a_i e^{-\text{Im}(\Omega_{i+})t} \left\{ \text{Re}(\vec{V}_i) \sin\left[\text{Re}(\Omega_{i+})t + \phi_{ia}\right] + \text{Im}(\vec{V}_i)\cos\left[\text{Re}(\Omega_{i+})t + \phi_{ia}\right] \right\} \quad (\text{I}.13)$$

$$\Delta \vec{x}_{i-} = b_i e^{-\text{Im}(\Omega_{i-})t} \left\{ \text{Re}(\vec{V}_i) \sin\left[\text{Re}(\Omega_{i-})t + \phi_{ib}\right] + \text{Im}(\vec{V}_i)\cos\left[\text{Re}(\Omega_{i-})t + \phi_{ib}\right] \right\} \quad (\text{I}.14)$$

where $\{a_i, b_i, \phi_{ia}, \phi_{ib}\}$ are unknown constants to be determined from initial conditions,

$$\text{Re}(\Omega_{i\pm}) = \frac{\mp \left[(\gamma^2 + 4m\,\text{Re}\{K_i\})^2 + 16m^2\,\text{Im}\{K_i\}^2\right]^{1/4} \sin(\delta_i/2)}{2m}$$

$$\text{Im}(\Omega_{i\pm}) = \frac{\gamma \pm \left[(\gamma^2 + 4m\,\text{Re}\{K_i\})^2 + 16m^2\,\text{Im}\{K_i\}^2\right]^{1/4} \cos(\delta_i/2)}{2m} \quad (\text{I}.15)$$

and

$$\delta_i = \begin{cases} \tan^{-1}\dfrac{4m\left|\text{Im}\{K_i\}\right|}{\gamma^2 + 4m\,\text{Re}\{K_i\}} & \text{if } \gamma^2 > 4m\,\text{Re}\{K_i\} \\ \pi - \tan^{-1}\dfrac{4m\left|\text{Im}\{K_i\}\right|}{\left|\gamma^2 + 4m\,\text{Re}\{K_i\}\right|} & \text{if } \gamma^2 < 4m\,\text{Re}\{K_i\} \end{cases} \quad (\text{I}.16)$$

**a) Complex unstable mode**

If Re{$K_i$}>0, $\Delta \vec{x}_{i+}(t)$ is spiraling inward to the equilibrium, whereas $\Delta \vec{x}_{i-}(t)$ is spiraling outward and its displacement diverges with time. Consequently, an optically trapped particle having a complex $K_i$ with positive real part is unstable and we denote this kind of solution as complex unstable mode.

**b) Quasi-stable mode**

If Re{$K_i$}<0, $\Delta \vec{x}_{i+}(t)$ is spiraling inward to the equilibrium. Here $\Delta \vec{x}_{i-}(t)$ requires some attention. The mode is spiraling outward if

$$\gamma < \gamma_{critical} = \frac{\sqrt{m}\,|\text{Im}(K_i)|}{\sqrt{|\text{Re}(K_i)|}}, \tag{I.17}$$

but spiraling inward if $\gamma > \gamma_{critical}$.

We denote this kind of solution as quasi-stable, where the stability depends on the damping provided by the environment. We note that the point $\gamma = \gamma_{critical}$ is non-hyperbolic, which simply means the linear term of the equation of motion vanishes, and the higher order terms are needed. As discussed in the main text, linear stability analysis is not sufficient to determine the stability at $\gamma \approx \gamma_{critical}$. Consequently, real time dynamics simulations are performed and the results are presented in Fig. 2 of the main text.

# Appendix II: The eigen force constant for various types of trapping beams

**A. The general form of force constant matrix and eigen force constants**

For an incident trapping beam, the general form of the force constant matrix for the trapped particle is

$$\ddot{\vec{K}} = \begin{bmatrix} a & d & 0 \\ g & b & 0 \\ e & f & c \end{bmatrix}, \tag{II.1}$$

where $a$, $b$, $c$, $d$, $e$, $f$, and $g$ are real numbers, and $K_{ij} = \partial(\vec{f}_{light})_i / \partial x_j$. Two of the components in Eq. (II.1), $K_{xz}$ and $K_{yz}$, are zero, because there is no induced force along the transverse plane as the particle is displaced along the $z$ axis. Here, we assume that the optical system including the focusing lens does not change the axial symmetry of the beam. By diagonalizing $\ddot{\vec{K}}$, we obtained the eigen force constants:

$$\begin{aligned} K_{axial} &= c, \\ K_{trans\pm} &= \left[ a + b \pm \sqrt{(a-b)^2 - 4dg} \right] / 2. \end{aligned} \tag{II.2}$$

When

$$-4dg < (a-b)^2, \tag{II.3}$$

the eigen force constants are all real numbers, and thus the nature of the optical trapping by such beam will be qualitatively similar to that of the conventional optical tweezers, i.e. all the eigen vibrational modes are stable modes (see Appendix I). However, when

$$-4dg > (a-b)^2, \tag{II.4}$$

$K_{trans\pm}$ are conjugate pair of complex numbers, and thus they correspond to the complex unstable mode or quasi-stable mode (see Appendix I).

## B. Numerical computation of eigen force constants for a variety of trapping beams.

In this section, we present the numerically computed eigen force constants for a particle in water trapped by a variety of different incident trapping beams. The incident trapping beams include (1) a linear polarized Gaussian beam (Fig. II.1), (2) a circularly polarized Gaussian beam (Fig. II.2), (3) a linear polarized Laguerre-Gaussian beam (Fig. II.3), (4) a right circularly polarized Laguerre-Gaussian beam (Fig. II.4), and (5) a left circularly polarized Laguerre-Gaussian beam (Fig. II.5).

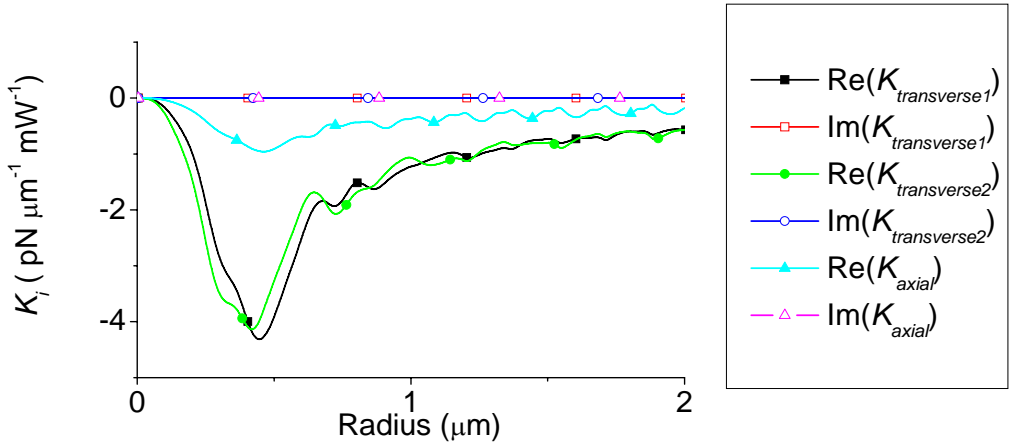

Fig. II.1. The eigen force constants for a particle ($\varepsilon_{sphere} = 1.57^2$) trapped by a linear polarized Gaussian beam with $f$=1, and N.A.= 1.2 in water ($\varepsilon_{water} = 1.33^2$).

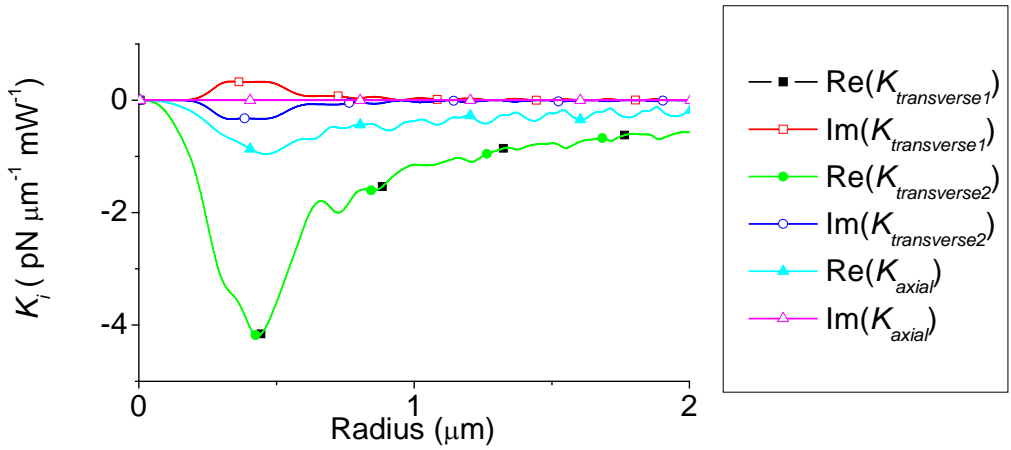

Fig. II.2. The eigen force constants for a particle ($\varepsilon_{sphere} = 1.57^2$) trapped by a circularly polarized Gaussian beam with $f=1$, and N.A.= 1.2 in water ($\varepsilon_{water} = 1.33^2$).

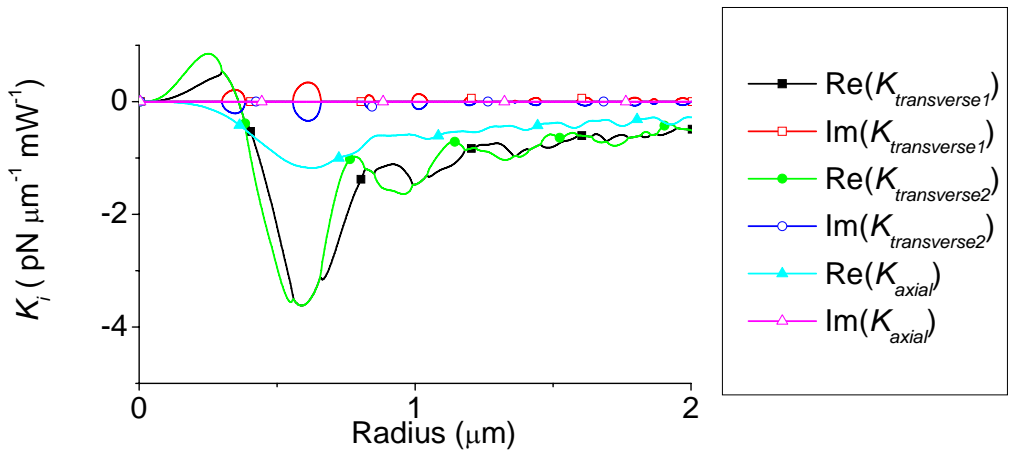

Fig. II.3. The eigen force constants for a particle ($\varepsilon_{sphere} = 1.57^2$) trapped by a linear polarized Laguerre-Gaussian beam with $l=1$, $f=1$, and N.A.= 1.2 in water ($\varepsilon_{water} = 1.33^2$).

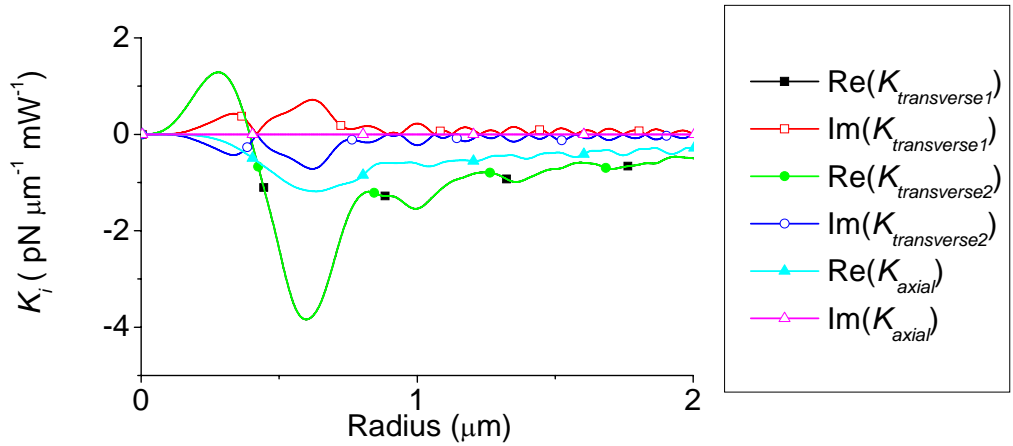

Fig. II.4. The eigen force constant for a particle ($\varepsilon_{sphere} = 1.57^2$) trapped by a right circularly polarized Laguerre-Gaussian beam with $l=1$, $f=1$, and N.A.= 1.2 in water ($\varepsilon_{water} = 1.33^2$).

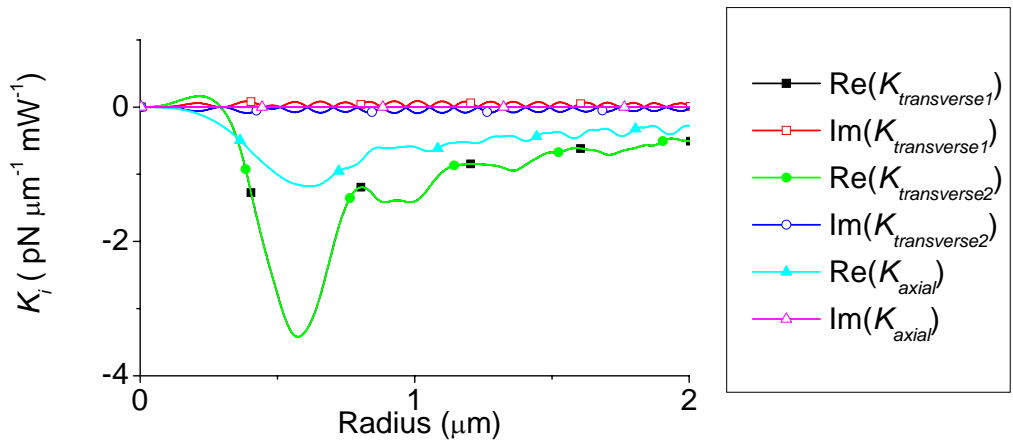

Fig. II.5. The eigen force constants for a particle ($\varepsilon_{sphere} = 1.57^2$) trapped by a left circularly polarized Laguerre-Gaussian beam with $l=1$, $f=1$, and N.A.= 1.2 in water ($\varepsilon_{water} = 1.33^2$).

## C. Trapping beams that carry no angular momentum

For an incident trapping beam that carries no angular moment, $d = g = 0$ as there is no rotating energy flux on the transverse plane (see main text). Accordingly, Eq. (II.4) can never be fulfilled, and thus the eigen force constants are always real numbers. From another perspective, the complex eigen force constants can only occur when the particle is allowed to exchange its angular momentum with the beam (see main text). Since the beam carries no angular momentum, complex eigen force constant should not occur.

The force constant matrix reduces to

$$\ddot{\vec{K}} = \begin{bmatrix} a & 0 & 0 \\ 0 & b & 0 \\ e & f & c \end{bmatrix}, \tag{II.5}$$

and the corresponding eigen force constants are

$$\begin{aligned} K_{axial} &= c, \\ K_{transverse1} &= a, \\ K_{transverse2} &= b, \end{aligned} \tag{II.6}$$

which are indeed real. The eigenvalues of a linearly polarized Gaussian beam, which carries no angular momentum, are plotted in Fig. II.1. Clearly, its eigenvalues are real numbers. The nature of optical trapping by a beam that carries no angular momentum will be qualitatively similar to that of the conventional optical tweezers.

**D. Cylindrically symmetric trapping beams that carry angular momentum**

A cylindrically symmetric optical vortex beam propagates along a helical path, which can drive the trapped particle to rotate (so $d = -g \neq 0$). Moreover, owing to the cylindrical symmetry, $a = b$. As such, the condition Eq. (II.4) is always fulfilled. Consequently, the corresponding transverse eigen force constants are always a conjugate pair of complex numbers.

To show this explicitly, consider a cylindrically symmetric optical vortex, such as a circularly polarized beam (Gaussian or Laguerre-Gaussian). It can be shown that $a = b$, as the restoring force acting on the particle when it is displaced along the $x$ axis is equal to that of the $y$ axis. Moreover, $d = -g$ because the induced torque when the particle is displaced along the $x$ axis is equal to that of the $y$ axis. Finally, $e = f$ because the induced force along the $z$ axis when the particle is displaced along the $x$ axis is equal to that of the $y$ axis. Substituting these expressions into (II.1), we obtain

$$\vec{\vec{K}} = \begin{bmatrix} a & d & 0 \\ -d & a & 0 \\ e & e & c \end{bmatrix}, \tag{II.7}$$

and the corresponding eigen force constants are

$$\begin{aligned} K_{axial} &= c, \\ K_{trans\pm} &= a \pm id. \end{aligned} \tag{II.8}$$

From Eq. (II.8), we see that complex eigen force constants occur whenever $d \neq 0$, as for any angular momentum carrying beam. In fact $d \neq 0$ indicates that there are angular momentum exchange between the beam and the particle, because complex eigenvalues can exist only when the trapped particle can exchange angular momentum with the beam (see main text). It is clear from (II.8) that the equilibrium cannot be solely characterized by real optical force constants. Loosely speaking, a particle in a cylindrically symmetric optical vortex can be considered as simultaneously experiencing a radial restoring force characterized by $\text{Re}(K_{trans}) = a$ and a torque about the beam's axis characterized by $\text{Im}(K_{trans}) = d$.

Fig. II.2, Fig. II.4, and Fig. II.5 show $K_i$ versus the radius of the trapped sphere, for a circularly polarized Gaussian beam, a right circularly polarized Laguerre-Gaussian beam, and a left circularly polarized Laguerre-Gaussian beam, respectively. Before entering the objective lens, the non-focused circularly polarized Gaussian beam carries spin angular momentum due to its polarization state, but not orbital angular momentum. After the beam is being strongly focused by the objective lens, part of its spin angular momentum is converted to orbital angular momentum (see main text). The left and right circularly polarized Laguerre-Gaussian beams have both spin and orbital angular momentum, in the former (later) case, the two forms of angular momentum are in opposite (same) direction. After focusing, the spin angular momentum is partially converted to orbital angular momentum. In the left (right) circular polarization case, the resultant angular momentum is small (large), owing to the cancellation (reinforcement) between the spin and orbital angular momentum. Consequently, $\text{Im}\{K_{trans}\}$ is the largest (smallest) for the right (left) circularly polarized Laguerre-Gaussian beam in general, because the spin and orbital angular momentum are reinforcing (cancelling) each others.

For all three beams, $K_{axial}$'s are always real and negative, indicating that the particle can be trapped along the axial direction due to gradient forces. On the contrary, $K_{trans\pm}$ are a conjugate pairs of complex numbers. For the Gaussian beam, $\text{Re}\{K_{trans\pm}\} < 0$, indicating that the particle can always be stabilized by introducing sufficient damping. On the other hand, for the Laguerre-Gaussian beams, $\text{Re}\{K_{trans\pm}\} > 0$ for particles that are smaller than the intensity ring of the beam, which means that small dielectric particles are unstable, as reported in experiments. Small dielectric particles are attracted toward intensity maxima. Under sufficient damping, these small particles will be orbiting along the high intensity ring of the beam. On the other hand, $\text{Re}\{K_{trans\pm}\} < 0$ for large particle. The spheres are bigger than the intensity ring, so that the gradient force drives the spheres to the beam center. A phase diagram is given in Fig. II.6, for a right circularly polarized LG beam with wavelength $\lambda = 1064$ nm, topological charge $l=1$, numerical aperture N.A.=1.2, and filling factor $f=1$. The trapped sphere is in water ($\varepsilon_{water} = 1.33^2$) and it has dielectric constant $\varepsilon_{sphere} = 1.57^2$ and mass density $\rho = 1050$ kg m$^{-3}$. At radius $R = 0.39\,\mu m$, $\gamma_{critical} \to \infty$ as $\text{Re}\{K_i\} \to 0$. The equilibrium point at $(x, y) = (0,0)$ is unstable for $R < 0.39\,\mu m$ for any values of damping. The particle will be trapped in the ring of the beam instead. For $R > 0.39\,\mu m$, the white region ($\gamma > \gamma_{critical}$) is the regime where sufficient damping can stabilize the particle, and the shaded region ($\gamma < \gamma_{critical}$) is where the damping is insufficient to stabilize the particle.

Compare Fig. II.6 with Fig. 2(b) of the main text, there two major differences. Firstly, $\gamma_{critical}$ is always greater than zero for the right circular polarization, whereas for the linear polarization, $\gamma_{critical}$ can be zero for some particle sizes. This is because the linear polarization does not possess cylindrical symmetric, therefore the condition Eq. (II.4) cannot always be fulfilled. Secondly, in general, the magnitude of $\gamma_{critical}$ for the right polarization is greater than that of the linear polarization. This is because the angular momentum of the right circularly polarized beam comes from both the spin and orbital angular momentum that are reinforcing each other, whereas that of the linearly polarized beam comes from the orbital angular momentum only. In both polarization, the envelope for $\gamma_{critical}$ increases linearly for large particle in general (see the blue dotted line in Fig. II.6). This is because the envelope of the force, and thus that of the eigen force constants, are proportional to $R^2$ (i.e. proportional to the geometrical cross section) for large particle. Then, according to (I.17), the envelope of $\gamma_{critical}$ increases linearly.

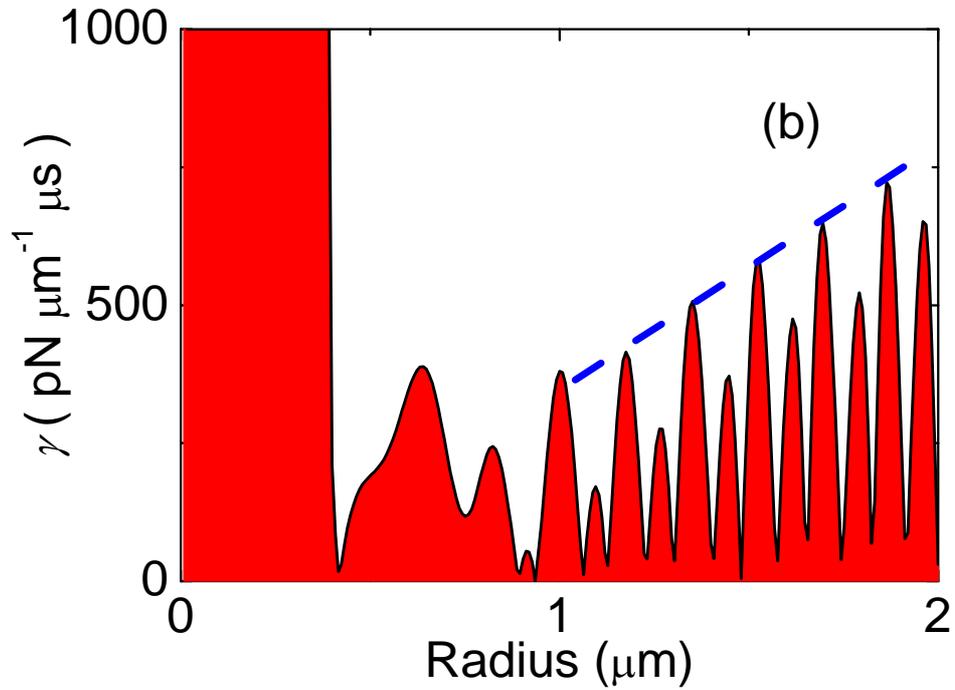

Fig. II.6 Phase diagram for a particle trapped at a power of 1W. The white (shaded) regions are unstable (stable). The black line is $\gamma_{critical}$. The incident beam is a right circularly polarized LG beam with $\lambda = 1064$ nm, $l=1$, $f=1$, and N.A.= 1.2.